\documentstyle[12pt,epsf]{article}
\begin{document}
\def\deg{^\circ}
\newcommand{\abseta}{\mid \eta \mid \leq}
\newcommand{\ptran}{{\rm P}_{\scriptscriptstyle\rm T}}

\begin{tabular}{lr}
ICHEP94 REF. GLS0122 & \hspace{1.75in}FERMILAB-CONF-94/129-E \\
Submitted to Pa 02, 17, 14 & \today \\
\hspace{1.0in}Pl 15, 19, 16 & \\
\end{tabular}

\vspace{1.5in}
\begin{center}
{\large \bf Measurement of correlated $b$ quark cross sections at CDF} \\
The CDF Collaboration \\
\end{center}

\begin{abstract}
Using data collected during the 1992-1993 collider run at Fermilab,  CDF has
made measurements of correlated $b$ quark cross sections where one $b$ is
detected from the lepton from semileptonic decay and the second $b$ is detected
with secondary vertex techniques. We report on measurements of the cross
section as a function of the momentum of the second $b$ and as a function of
the azimuthal separation of the two $b$ quarks, for transverse momentum of the
initial $b$ quark greater than 15 GeV.  The vertex reconstruction techniques
are valid over a large range in transverse momentum, starting at a minimum of
10 GeV.  Results are compared to QCD predictions.
\end{abstract}

\vspace{3.0in}
\begin{flushleft}
Contact Person:  Dr. Paul F. Derwent\\
University of Michigan\\
E-Mail:  derwent@fnald.fnal.gov\\
\end{flushleft}

\newpage

\par Studies of $b$ production in $p{\overline p}$ collisions provide
quantitative tests of perturbative QCD.  For processes involving
momentum transfers on the order of $m_b$, the strong coupling constant,
$\alpha_s$, becomes relatively small and perturbative methods are expected to
work well.  Measurements of inclusive cross sections for $p{\overline
p}\rightarrow bX$ have been made at CDF~\cite{latest_CDF} and
UA1~\cite{old_UA1}.
Consideration of the process $p{\overline p}\rightarrow
b{\overline b}X$ provides further opportunities for comparison of experiment
and QCD calculations.

\par We will make measurements of correlated $b$ quark cross sections,
where one $b$ is detected
from the lepton from semileptonic decay and the second $b$ is detected with
secondary vertex techniques.  We report on measurements of the
cross section as a function of the transverse momentum of the second $b$
(d$\sigma_b$/dE$_T$) and as a function of the azimuthal
separation (d$\sigma$/d$\delta\phi$) of the two $b$ quarks,
for transverse momentum of the initial $b$ quark greater than 15 GeV.

\par This paper presents the first correlated cross sections measured at CDF
using a combination of lepton and vertex tags to identify $b$ events.  Starting
with a well identified $\mu$ candidate,  we look for the presence of a second
$b$ with secondary vertex techniques.  Using jets with E$_T>$ 10 GeV, we
measure the cross section as a function of the ${\overline b}$ transverse
momentum and also the cross section as a function of the azimuthal angle
between the $b$ and ${\overline b}$ quarks.  For purposes of notation, we will
consider the $\mu$ as coming from a $b$ quark, and the jet as coming from a
${\overline b}$ quark.

\par The CDF has been described in detail elsewhere~\cite{NIM_book}.  The
tracking
systems used for this analysis are the silicon vertex detector (SVX), the
central tracking chamber (CTC), and the muon system.  The SVX and CTC are
located in a 1.4 T solenoidal magnetic field.  The SVX consists of 4 layers of
silicon-strip detectors with $r-\phi$ readout, including pulse height
information~\cite{SVX_paper}, with a total active length of 51 cm.  The pitch
between readout strips is 60 $\mu$m and a spatial resolution of 13 $\mu$m has
been obtained.  The first measurement plane is located 2.9 cm from the
interaction point, leading to an impact parameter resolution of $\approx$ 15
$\mu$m for tracks with transverse momentum, p$_t$, greater than 5 GeV/c.  The
CTC is a cylindrical drift chamber containing 84 layers, which are grouped into
alternating axial and stereo superlayers containing 12 and 6 wires
respectively, covering the radial range from 28 cm to 132 cm.  The central
muon system consists of two detector elements.  The Central Muon chambers
(CMU), located behind $\approx$ 5 absortion lengths of material, provide muon
identification over 85\% of $\phi$ for the pseudorapidity range
$|\eta|\leq$0.6, where $\eta = - \ln[\tan(\theta/2)]$.  This $\eta$ region is
further instrumented by the Central Muon Upgrade chambers (CMP), located after
$\approx$
8 absorption lengths.  The calorimeter systems used for this analysis are the
central and plug systems.  The central subtends the range $|\eta|< $ 1.1 and
spans 2$\pi$ in azimuthal coverage.  The plug subtends the range $1.1 < |\eta|
< 2.4$, again with 2$\pi$ azimuthal coverage.

\par CDF uses a three-level trigger system.  At Level 1, muon candidate events
are selected with a trigger that requires the presence of a hit pattern,
consistent with p$_t>$ 6 GeV/c, in the CMU chambers and confirming hits in the
CMP chambers.  At Level 2, the trigger
requires that the CMU chamber track match a track found in the CTC, using the
Central Fast Track processor, with p$_t>$ 9.2 GeV/c.  At Level 3, the trigger
requires a muon track in both the CMU and CMP chambers matched with a track
found in the CTC, using the offline
track reconstruction algorithm, with p$_t>$ 7.5 GeV/c.

\par An inclusive muon sample is formed from the muon triggered sample with the
following selection criteria:  (1)  reconstructed CTC track, p$_t > $ 9 GeV/c,
(2) muon tracks exist in both the CMU chambers and CMP chambers, (3) the
$\chi^2$ of the match betweent muon
track and the extrapolated CTC track match be less than 9 in the transverse
view of both CMU and CMP chambers and be less than 12 for the longitudinal view
in the CMU chambers, and (4) the track extrapolate and is found
in the SVX fiducial region.  There are 145784 events passing all requirements
in this data sample.

\par Monte Carlo samples for $b$ and $c$ quarks are produced using ISAJET
version 6.43~\cite{isajet}.  The CLEO Monte Carlo program~\cite{qq} is used to
model the decay of $b$ hadrons.  $b$ quarks produced using the HERWIG Monte
Carlo~\cite{herwig} are also used for systematic studies.  One of the final
state $b$ partons is required to have p$_t >$ 15 GeV. Events with a $\mu$ with
p$_t >$ 8 GeV are passed through the full CDF simulation and reconstruction
package.  The simulation used an average $b$ lifetime of $c\tau$ = 420 $\mu$m.

\par The $\mu$ acceptance and efficiency has three parts to it: (1) the
fiducial acceptance for muons coming from $b$'s with $|\eta|<1$, (2) the
fraction of $b$'s, $p_t^b > p_t^{min}$, which decay to muons with p$_t>$ 9 GeV
and (3) the trigger and identification efficiencies for 9 GeV muons. The first
two factors have been studied using the Monte Carlo samples described above.
The trigger and identification efficiencies for muons are studied from the
data, using J/$\psi$ and Z$^{\circ}$ samples.  All acceptance and efficiency
numbers are summarized in table~\ref{table-mu_effs}.

\par The $p_t^{min}$ value is chosen according to the standard interpretation,
where 90\% of muons with p$_t >$ 9 GeV come from $b$ quarks with $p_t$ greater
than $p_t^{min}$.  For this sample, the $p_t^{min}$ value is 15 GeV, from
studying a sample of $b$ quarks generated with $p_t >$ 10 GeV.

\par  We find that 18.6 $\pm$ 1\% of  muons with p$_t>$ 9 GeV pass through the
CMU---CMP fiducial regions, where the error is statistical only. This value is
for the entire $\eta$ distribution of produced $b$ quarks.  Restricting the
$|\eta|$ of the parent $b$ quark to be less than 1, we find 39.2 $\pm$ 2\% of
muons with p$_t>$ 9 GeV pass through the CMU---CMP chambers.   From a study of
$b$ quarks with $p_t >$ 15 GeV which decay to $\mu$, we find that 10.7 $\pm$
0.1 \% (statistical error only) have p$_t >$ 9 GeV.  We use the branching ratio
B($b\rightarrow\mu$) = 0.108 $\pm$ 0.065 which comes from direct measurements
at CLEO~\cite{CLEO_br}.  This  fraction includes the sequential decay
contribution, which has been scaled by the relative branching ratios of
$b\rightarrow\mu$ to $c\rightarrow\mu$.  We have varied the mean $<z>$ used in
the $b$ fragmentation~\cite{fragmentation} to study the effects in the
geometric and kinematic acceptance.  The systematic errors are therefore
correlated.  A 1$\sigma$ variation in the mean $<z>$ gives a $+1~-6.5$ \%
change in the geometric acceptance and a $+9.7~-10.8$\% change in the kinematic
acceptance.

\par The trigger efficiency is measured using independently triggered samples
for each level of the system.  The efficiency curves are then convoluted  with
the p$_t$ spectrum of muons, to get the efficiency for a muon with p$_t >$ 9
GeV.  The combined efficiency of the L1, L2, and L3 triggers is measured  to be
0.81 $\pm$ 0.024.   The efficiency of the muon reconstruction algorithms has
been studied in detail in the muon plus charm meson cross section
work~\cite{t_lecompte} and  is found to be 0.981 $\pm$ 0.003.  Given the
presence of reconstructed stubs in the chambers, the matching efficiency has
been studied in a J/$\psi$ sample and found to be 0.987 $\pm$ 0.013 for the
matching cuts used in this analysis.   The requirement that the CTC track be in
the SVX fiducial region, in combination with the SVX tracking efficiency, has
an efficiency of 0.68 $\pm$ 0.021.

\begin{table}
\begin{center}
\begin{tabular}{|c|c|}       \hline
Geometric Acceptance & 0.392 $+0.005~-0.026$ \\ 
Kinematic Acceptance & 0.107 $+0.010~-0.012$ \\ 
Branching Ratio & 0.108 $\pm$ 0.065 \\ \hline	
Trigger Efficiency & 0.81 $\pm$ 0.024 \\ 	
Reconstruction Efficiency & 0.981 $\pm$ 0.003 \\
Matching Efficiency & 0.987 $\pm$ 0.013 \\	
SVX requirement & 0.68 $\pm$ 0.021 \\ \hline	
Combined Acceptance  & \\
and Efficiency & 0.00239 $+0.00030~-0.00018$ \\ \hline 
\end{tabular}
\end{center}
\caption{Summary of muon acceptance and efficiency numbers.  The uncertainties
in the kinematic and geometric acceptance are correlated.}
\label{table-mu_effs}
\end{table}

\par The combined acceptance and efficiency for muons coming from $b$ quarks
with $p_t > $ 15 GeV is 0.00239 $+0.00030~-0.00018$. This correction will be
applied to all cross section numbers presented below. Therefore, there is a
common uncertainty of $+12.4~-7.5$\% on all the cross sections coming from the
$\mu$ acceptance and identification efficiency.

For the correlated cross sections, we look for additional jets in the sample,
binning by jet E$_T$ and azimuthal opening angle.  For each jet in the sample,
we calculate a probability that the jet is consistent with coming from the
primary vertex.  Using input templates from $b$, $c$, and primary (light quark
and gluon) jet Monte Carlo samples, we fit the data as a sum of these three
processes. Assuming that the presence of the $\mu$ candidate signals an
independent  $b$ quark, we can then make correlated cross section measurements.
Correcting for the $\mu$ acceptance and identification requirements  and the
${\overline b}$ jet acceptance, and scaling by the integrated luminosity of
the sample, we have differential cross section measurements.

\par We make use of a jet probability algorithm~\cite{PRD} which asks the
question ``Is the ensemble of  tracks in this jet consistent with being from
the primary vertex?''   This algorithm compares track impact parameters to
measured resolution functions in order to calculate for each jet a probability
that it is attached to the origin. This probability is uniformly distributed
for light quark or gluon jets, but is very low for jets with displaced vertices
from heavy flavor decay.  The algorithm assigns a probability between 0 and 1,
where a  probability value near 1 means the jet is very consistent with being
primary, and a probability value near 0 means the jet is very inconsistent with
being primary.

\par For jet clusters identified in the calorimeter, the algorithm selects a
set of tracks, p$_t >$ 1 GeV/c,  within a cone of 0.4 around the jet axis to be
used in the calculation of the jet probability.  There are loose cuts to
affiliate the tracks with the primary vertex, in addition to track quality
requirements~\cite{PRD}.  We require that there be $\geq$ 2 tracks passing the
quality requirements for the calculation of the probability.

\par Jets in this sample are required to have E$_T>$ 10 GeV in a cone of radius
0.4, $|\eta|$ $<$ 1.5,
have at least 2 good tracks, and be separated from the muon in
$\eta-\phi$ space by $\Delta$R $\geq$ 1.0.  The cone of radius 1.0 was
chosen so that the tracks clustered around the jet axis were separated from the
$\mu$ direction.  All jet energies in this paper are measured energies, not
including corrections for known detector effects~\cite{jet_papers}.

\par  The ${\overline b}$ jet acceptance combines the fiducial acceptance of
the SVX and the CTC, the track reconstruction efficiency, and fragmentation
effects.  The simulation is used to calculate the combination of these  pieces.
The acceptance represents the fraction of $b$ quarks which produce jets with
E$_T >$ 10 GeV, $|\eta| <$ 1.5 and at least 2 good tracks inside a cone of 0.4
around the jet axis, where there is also a $b$ quark which decays to a $\mu$
with p$_t >$ 9 GeV within the CMU-CMP acceptance.   The ${\overline b}$ jet
acceptance is calculated separately as a function of the jet E$_T$ and
azimuthal opening angle between the two quarks.

\par The average acceptance for the ${\overline b}$ is $\approx$~40\%.  It
ranges from 32.9 $\pm$ 1.9\% (statistical error only) for 10 $<$ E$_T$ $<$ 15
GeV to 49.8 $\pm$ 7.3\% for 40 $<$ E$_T$  $<$ 50 GeV.  For $\delta\phi <
22.5\deg$, the acceptance is 7.3 $\pm$ 2.2\%, while for
157.5$\deg < \delta\phi < 180\deg$, the acceptance is 51.4 $\pm$ 0.8\%.
Tables~\ref{table-et_acceptance} and \ref{table-phi_acceptance} show the bin by
bin values used in the differential cross section measurements.

\begin{table}
\begin{center}
\begin{tabular}{|c|c|}       \hline
E$_T$ Range & Acceptance \\ \hline\hline
10 --- 15 & 32.9 $\pm$ 1.9 \% \\
15 --- 20 & 46.3 $\pm$ 2.2 \% \\
20 --- 25 & 48.8 $\pm$ 2.9 \% \\
25 --- 30 & 50.8 $\pm$ 3.9 \% \\
30 --- 40 & 49.7 $\pm$ 4.2 \% \\
40 --- 50 & 49.8 $\pm$ 7.3 \% \\ \hline
\end{tabular}
\end{center}
\caption{${\overline b}$ jet acceptance as a function of jet E$_T$ for $|\eta|
<$ 1.5  (statistical errors only).}
\label{table-et_acceptance}
\end{table}

\par  We have compared the values for the ${\overline b}$ jet acceptance from
ISAJET samples to the acceptance from HERWIG samples.  The acceptance agrees
within the statistical error in the samples as a function of E$_T$, differing
at the 5\% level.  We will take this as an additional systematic uncertainty on
the acceptance.  In combination with a 10\% uncertainty due to the vertex
distribution for events in the SVX fiducial volume, we have a common  11.2\%
systematic uncertainty in all the jet acceptance numbers.

\begin{table}
\begin{center}
\begin{tabular}{|c|c|}       \hline
$\delta\phi$ Range & Acceptance \\ \hline\hline
0 --- 22.5$\deg$ & 7.25 $\pm$ 2.2 \% \\
22.5$\deg$ --- 45$\deg$ & 4.52 $\pm$ 2.0 \% \\
45$\deg$ --- 67.5$\deg$ & 36.9 $\pm$ 3.0 \% \\
67.5$\deg$ --- 90$\deg$ & 51.4 $\pm$ 0.8 \% \\
90$\deg$ --- 112.5$\deg$ & 51.4 $\pm$ 0.8 \% \\
112.5$\deg$ --- 135$\deg$ & 51.4 $\pm$ 0.8 \% \\
135$\deg$ --- 157.5$\deg$ & 51.4 $\pm$ 0.8 \% \\
157.5$\deg$ --- 180$\deg$ & 51.4 $\pm$  0.8 \% \\ \hline
\end{tabular}
\end{center}
\caption{${\overline b}$ jet acceptance as a function of $\delta\phi$ of the
two $b$ quarks, for E$_T >$ 10 and $|\eta| <$ 1.5 (statistical errors only).
The first three bins have lower acceptance due to the $\Delta$R cut for
identification of the two $b$ jets.}
\label{table-phi_acceptance}
\end{table}



\par We use a binned maximum likelihood fit~\cite{Wprime} to distinguish the
$b$, $c$, and primary jet contributions in the sample.
In this case, we use the results of the jet
probability algorithm to define the variable that is used in the fit.  We find
that the log$_{10}$(jet probability) shows stronger differentiation between
$b$, $c$, and primary jets (see figure~\ref{fig-smoothed_shapes}) than the jet
probability and will use this variable in the fitting algorithm. We fit over
the  range -10 --- 0 in log$_{10}$(jet probability), where the $b$, $c$, and
primary contributions are constrained to be positive, but no other constraints
are included in the fit.  We use smoothed Monte Carlo distributions from $b$
and $c$ samples as the input shapes for heavy flavor to decrease the effects of
limited Monte Carlo statistics.  We model the primary jets with an exponential
distribution, since a logarithm transforms a uniform distribution to an
exponential distribution.

\par We have explored the effect of different Monte Carlo samples to form the
input shape used in the fit.  Using different input $b$ Monte Carlo samples
compared to a test distribution made with independent Monte Carlo samples
shows a 5\% change in the fit fractions.  Changing the average $b$ lifetime by
6\%~\cite{b_lifetime} changed the fit fraction by 3\%.  We include a 5.8\%
systematic uncertainty to our fit results to account for systematic errors in
the fitting procedure and uncertainty of the $b$ lifetime.

\par  In figure~\ref{fig-sample_fit}, we show the distribution of
log$_{10}$(jet probability) for all jets in the $\mu$ sample, overlayed with
the fit results.
In this sample, the fit predicts 2620 $\pm$ 97 $b$ jets, 2085
$\pm$ 180 $c$ jets, and 13103 $\pm$ 161 primary jets for a total of 17808.
There are 17810 events in the data sample.  Figure~\ref{fig-fit_comparisons}
shows three comparisons of the data and fit results, showing the bin-by-bin
difference in the results, the bin-by-bin difference divided by the errors, and
the distribution of the difference divided by the errors.  In these
distributions, the errors are a combination of the statistical errors in the
data points and the overall scale error in the fitted values.  We do not
include any error on the Monte Carlo shapes. From these
distributions, we can see that the inputs model the data well.  The difference
divided by the errors has a mean of -0.04 and RMS of 1.08.

\par We have applied the fitting algorithm to several different samples of
jets, where we have made different requirements on the transverse energy of the
jet or the azimuthal separation between the $\mu$ and the jet.  Applying the
acceptance and efficiency numbers for the $\mu$ candidate, the ${\overline b}$
jet acceptance and dividing by the integrated luminosity of the sample, we
convert the number of fit ${\overline b}$ jets into differential cross
sections. All of the cross section numbers presented have a set of
uncertainties in common.  These come from the $\mu$ acceptance and
identification ($+12.4~-7.5$\%), the fitting procedure (5.8\%), the vertex
acceptance (10\%), the ${\overline b}$ jet acceptance (5\%), and the luminosity
normalization (3.6\%).  The total common systematic error is $+18.3~-15.4\%.$

\par We look at the E$_T$ distribution of the jet in the event, using 6 bins to
cover the region between 10 GeV and 50 GeV in transverse energy.  In each E$_T$
bin, we do an independent fit of the log$_{10}$(jet probability) distribution
and then correct for the acceptance.  Table~\ref{table-et_cross_sections}
contains a summary of the number of ${\overline b}$ jets, number of total jets,
and the cross section in each E$_T$ bin considered.
Figure~\ref{fig-et_comparison} shows the values of the cross section as a
function of jet E$_T$ for $\mu$ with $>$ 9 GeV.

\begin{table}
\begin{center}
\begin{tabular}{|c|c|c|c|}
\hline
E$_T$ range & Number of Jets & Number of fit ${\overline b}$ jets &
Cross Section (nb/GeV)\\
\hline\hline
10 --- 15 & 5695 & 650 $\pm$ 52 & 5.49 $\pm$ 0.63 \\
15 --- 20 & 4083 & 647 $\pm$ 48 & 3.88 $\pm$ 0.41 \\
20 --- 25 & 2680 & 452 $\pm$ 40 & 2.57 $\pm$ 0.32 \\
25 --- 30 & 1774 & 323 $\pm$ 33 & 1.77 $\pm$ 0.25 \\
30 --- 40 & 1915 & 308 $\pm$ 32 & 0.86 $\pm$ 0.13 \\
40 --- 50 & 835 & 145 $\pm$ 21 & 0.41 $\pm$ 0.085 \\
\hline
\end{tabular}
\end{center}
\caption{Cross section summary as a function of jet E$_T$.  There is a common
systematic uncertainty of $+18.3~-15.4$\% not included in the experimental
points.}
\label{table-et_cross_sections}
\end{table}

\par  We also measure the cross section as a function of the azimuthal
separation between the jet and the $\mu$.  The measured jet direction is a good
measure of the initial ${\overline b}$ direction, while  the $\mu$ follows the
$b$ direction.  We have chosen to use bins of $22.5\deg$ in $\delta\phi$,
which is the width of the spread between the $b$ direction and the $\mu$
direction.  With this bin size, the azimuthal separation of the jet and $\mu$
is a good approximation of the azimuthal separation of the two quarks.
Table~\ref{table-phi_cross_sections} contains a summary of the number of jets,
number of fitted ${\overline b}$ jets, and the cross section in each
$\delta\phi$ bin considered.

\begin{table}
\begin{center}
\begin{tabular}{|c|c|c|c|}       \hline
$\delta\phi$ Range & Number of Jets & Number of fit ${\overline b}$ jets &
Cross Section (nb/$22.5\deg$)\\ \hline\hline
0 --- 22.5$\deg$ & 83 & 11.7 $\pm$ 7.4 & 2.24 $\pm$ 1.58 \\
22.5$\deg$ --- 45 $\deg$ & 143 & 35.8 $\pm$ 10.6 & 10.88 $\pm$ 5.78 \\
45$\deg$ --- 67.5$\deg$ & 364 & 73 $\pm$ 16 & 2.25 $\pm$ 0.66 \\
67.5$\deg$ --- 90$\deg$ & 442 & 81 $\pm$ 17 & 2.19 $\pm$ 0.48 \\
90$\deg$ --- 112.5$\deg$ & 624 & 122 $\pm$ 20 & 3.30 $\pm$ 0.58 \\
112.5$\deg$ --- 135$\deg$ & 1188 & 195 $\pm$ 27 & 5.27 $\pm$ 0.80 \\
135$\deg$ --- 157.5$\deg$ & 3394 & 477 $\pm$ 43 & 12.9 $\pm$ 1.40 \\
157.5$\deg$ --- 180$\deg$ & 11752 & 1624 $\pm$ 76 & 43.9 $\pm$ 3.34 \\ \hline
\end{tabular}
\end{center}
\caption{Cross section summary as a function of $\delta\phi$ between the jet
and $\mu$.  There is a common systematic uncertainty of $+18.3~-15.4$\% not
included in the experimental points.}
\label{table-phi_cross_sections}
\end{table}

\par  Figure~\ref{fig-phi_comparison} shows a comparison to a theoretical
prediction, using the Mangano-Nason-Ridolfi calculation~\cite{MNR} with $m_b$ =
4.75 GeV, MRSB structure functions, and $\mu^2$ = $m_b^2$ + $<p_t>^2$ for
$|\eta^{b}|<$ 1., $p_t^{b}>$ 15 GeV, $|\eta^{{\overline b}}|<$ 1.5, and
$p_t^{{\overline b}}>$ 10 GeV as input.   We have smeared the parton $p_t$
distribution with the expected energy scale and resolution of the CDF
detector~\cite{jet_production}, and have normalized the
$\delta\phi$ distribution to the cross section with smeared $p_t >$
10 GeV.  The shapes of the
theoretical prediction and the experimental data agree well, especially for
$\delta\phi > 90\deg$.  Note that there is a large change in the
acceptance for $\delta\phi < 60\deg$ (see
table~\ref{table-phi_acceptance}) due to the $\Delta$R separation requirement
on the $\mu$ jet system.

\par In conclusion, we have presented the first measurements of correlated
$b\overline{b}$ cross sections using a combination of lepton and vertexing
techniques to identify the two $b$'s.  We measure the differential transverse
energy cross section for the ${\overline b}$, d$\sigma_{{\overline
b}}/$dE$_T$, and the differential azimuthal cross section between
the two quarks,  d$\sigma/$d$\delta\phi$.  The shape in the
$\delta\phi$ distribution agrees
well with theoretical predictions, but the overall normalization is roughly a
factor of 1.3 higher than predicted.

\par We thank the Fermilab staff and the technical staffs of the
participating institutions for their vital contributions.  This work was
supported by the U.S. Department of Energy and National Science Foundation;
the Italian Istituto Nazionale di Fisica Nucleare; the Ministry of Education,
Science and Culture of Japan; the Natural Sciences and Engineering Research
Council of Canada; the National Science Council of the Republic of China;
the A. P. Sloan Foundation; and the Alexander von Humboldt-Stiftung.

%
%

\newpage

\begin{figure}
\epsfysize=7in
\epsffile[0 90 594 684]{smoothed_shapes.ps}
\caption{The log$_{10}$(jet probability) distributions used as inputs to the
fitting program.  The $b$ and $c$ shapes are smoothed versions of Monte Carlo
distributions, while the primary shape is an exponential function.  The three
distributions are normalized to equal area and shown on the same vertical
scale.}
\label{fig-smoothed_shapes}
\end{figure}

\begin{figure}
\epsfysize=5in
\epsffile[-90 90 504 684]{sample_fit.ps}
\caption{For 10 GeV jets in the $\mu$ sample, we show the data distribution
overlayed with the fit results.  Statistical errors on the data and scale
errors on the fit results are included.  The fit results model the data well
over the entire range of the fit.}
\label{fig-sample_fit}
\end{figure}

\begin{figure}
\epsfysize=7in
\epsffile[0 90 594 684]{fit_comparisons.ps}
\caption{Various comparisons of the data distribution and the fit results.  We
show the bin by bin difference between the data and the fit results, the
bin-by-bin difference scaled to the errors, and the distribution of the
difference scaled to the errors.  In all cases, the errors are the statistical
error in the data points plus the overall scale error in the fitted values.}
\label{fig-fit_comparisons}
\end{figure}

\begin{figure}
\epsfysize=7in
\epsffile[0 90 594 684]{et_comparison.ps}
\caption{The distribution of the second $b$ jet, E$_T > $ 10 GeV and $|\eta| <$
1.5, cross section as a function of jet E$_T$, given a $\mu$ with p$_t > $ 9
GeV present in the event.  There is a common systematic uncertainty of
$+18.3~-15.4$\% not included in the experimental points.}
\label{fig-et_comparison}
\end{figure}

\begin{figure}
\epsfysize=7in
\epsffile[0 90 594 684]{phi_comparison.ps}
\caption{The distribution of the cross section as a function of the azimuthal
angle between the $\mu$ and the jet, with E$_T >$ 10 GeV and $|\eta| <$ 1.5,
overlayed with a theoretical prediction.   The theoretical prediction has been
normalized to the expected cross section after including the effects of
detector smearing.   There is a common systematic uncertainty of
$+18.3~-15.4$\% not included in the experimental points.  Note that the
acceptance changes rapidly and has large uncertainty for the region $\delta\phi
<$ 60$\deg$ due to the effects of the $\Delta$R separation cut.}
\label{fig-phi_comparison}
\end{figure}

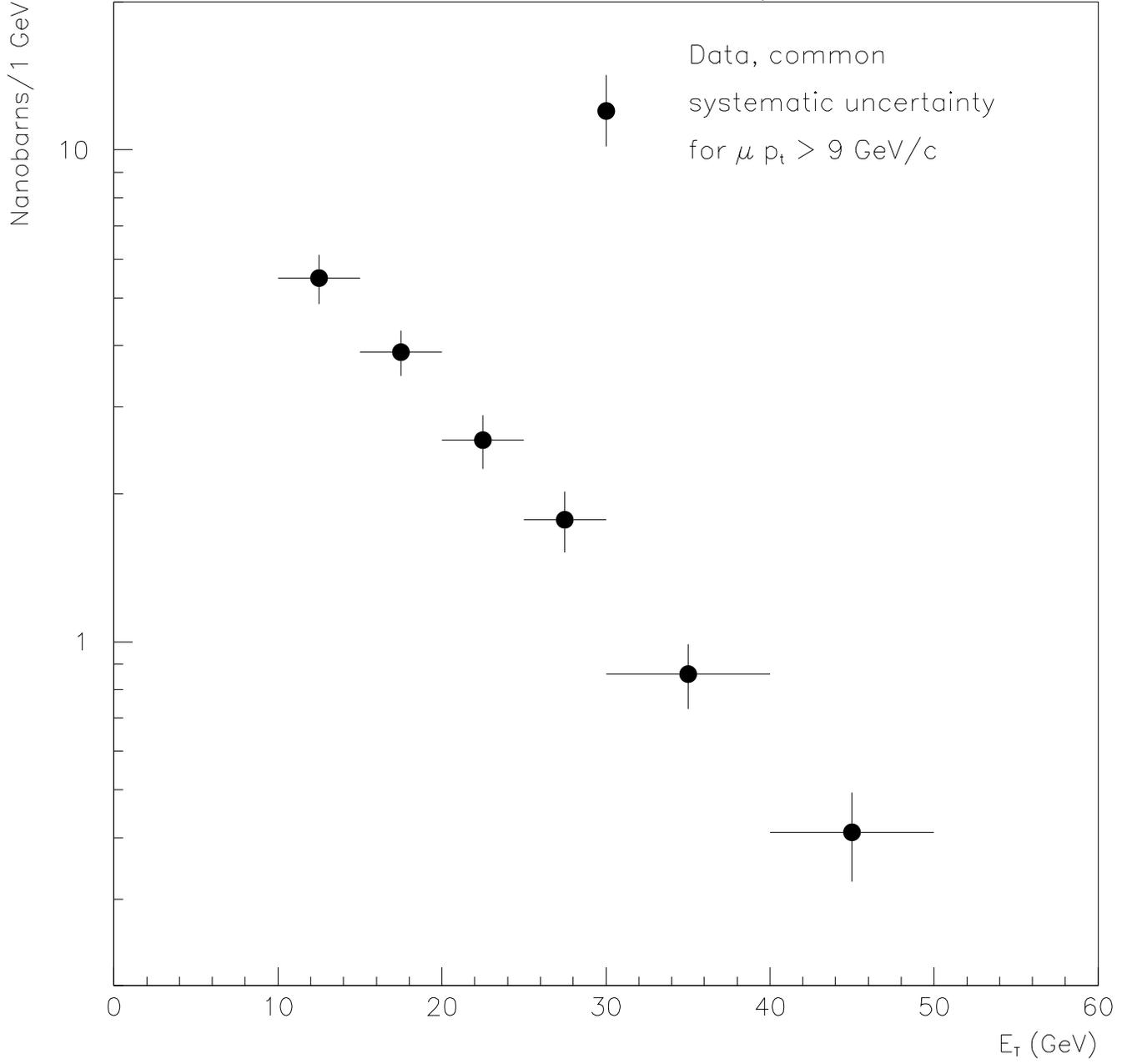
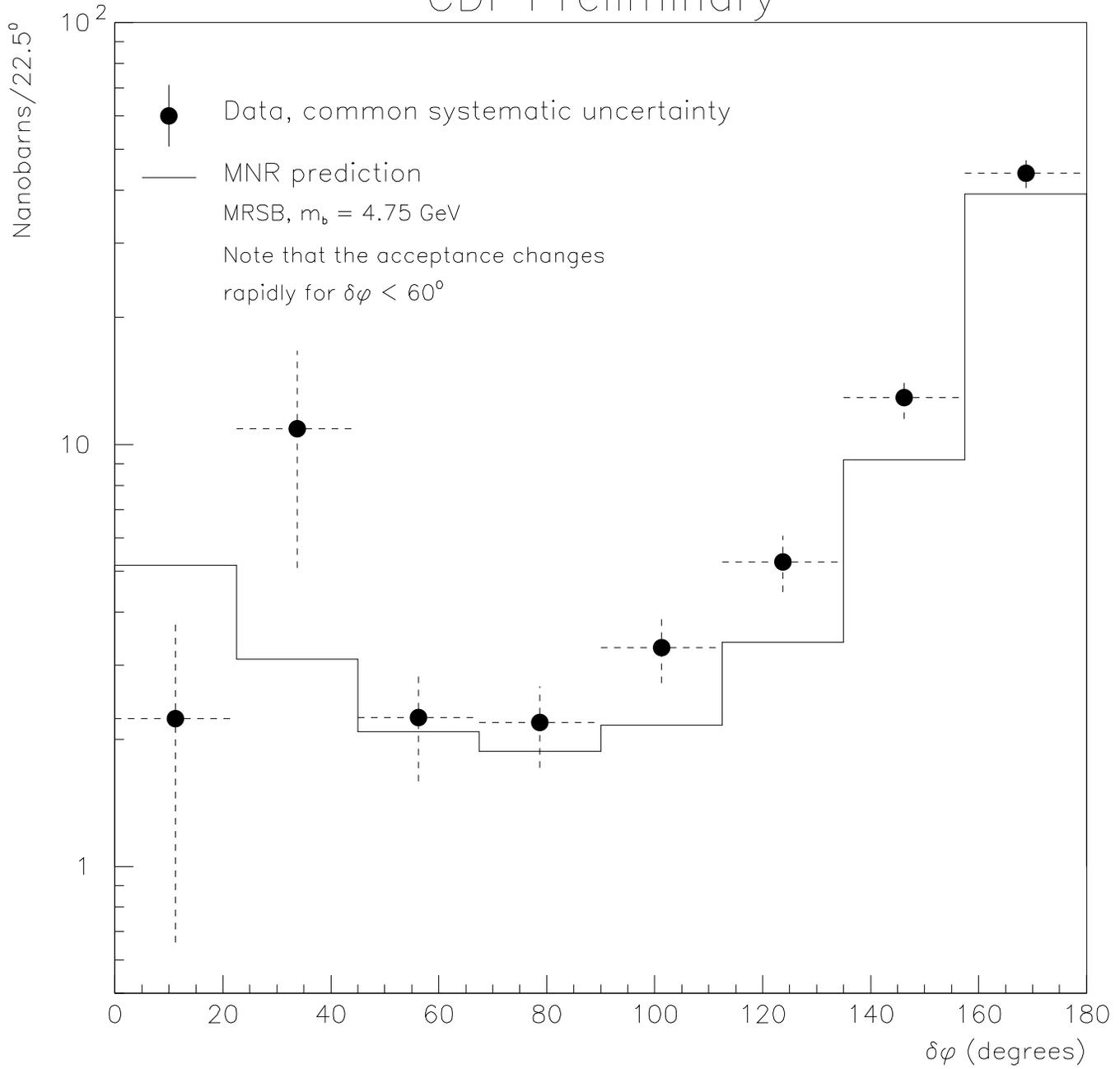
\end{document}